# Population Growth, Energy Use, and the Implications for the Search for Extraterrestrial Intelligence

B. Mullan[1,2] and J. Haqq-Misra[1]

**Abstract:**

Von Hoerner ("Population Explosion and Interstellar Expansion," Journal of the British Interplanetary Society, 28, 691-712, 1975; hereafter VH75) examined the effects of human population growth and predicted agricultural, environmental, and other problems from observed growth rate trends. Using straightforward calculations, VH75 predicted the "doomsday" years for these scenarios (≈2020–2050), when we as a species should run out of space or food, or induce catastrophic anthropogenic climate change through thermodynamically unavoidable direct heating of the planet. Now that over four decades have passed, in this paper we update VH75. We perform similar calculations as that work, with improved data and trends in population growth, food production, energy use, and climate change. For many of the impacts noted in VH75 our work amounts to pushing the "doomsday" horizon back to the 2300s-2400s (or much further for population-driven interstellar colonization). This is largely attributable to using worldwide data that exhibit smaller growth rates of population and energy use in the last few decades. While population-related catastrophes appear less likely than in VH75, our continued growth in energy use provides insight into possible future issues. We find that, if historic trends continue, direct heating of the Earth will be a substantial contributor to climate change by ≈2260, regardless of the energy source used, coincident with our transition to a Kardashev type-I civilization. We also determine that either an increase of Earth's global mean temperature of ≈12K will occur or an unreasonably high fraction of the planet will need to be covered by solar collectors by ~2400 to keep pace with our growth in energy use. We further discuss the implications in terms of interstellar expansion, the transition to type II and III civilizations, SETI, and the Fermi Paradox. We conclude that the "sustainability solution" to the Fermi Paradox is a compelling possibility.

---

[1] Blue Marble Space Institute of Science, 1001 4th Ave, Suite 3201 Seattle, WA 98154 (brendan@bmsis.org)
[2] Point Park University Department of Natural Sciences, Engineering, and Technology, 201 Wood Street, Pittsburgh PA 15222 (bmullan@pointpark.edu)

**Highlights:**

- We update calculations of population- and energy-related "doomsdays" from Von Hoerner (1975).
- Current trends show "doomsdays" related to a population singularity or agricultural limitations may not occur.
- "Doomsdays" may occur due to greenhouse gas emissions by ~2300, and/or direct heating by 2300–2400.
- Using ~$10^{16}$-$10^{17}$ W may raise Earth's temperature by 12K or require complete coverage by solar collectors by ~2400
- If this also applies to extraterrestrial civilizations, it may support the "sustainability solution" to the Fermi Paradox.

**1. Introduction**

Human populations, like other organisms, are subject to environmental factors that can ultimately place limits on their growth. The maximum population that can be sustained in a given environment is set by the environment's "carrying capacity," with populations that exceed this maximum forced into decline or collapse. Thomas Malthus (1789) famously postulated that humanity would return to subsistence levels in the future as a result of rapid population growth outpacing food production. However, Malthus may not have foreseen the subsequent developments in technology that would substantially increase the carrying capacity for human civilization, allowing the population to grow from just under a billion people when he wrote in 1789 to more than seven billion today. Human civilization is not immune from environmental limits to growth, but we do retain the capacity to alter our environment in order to improve the suitability of the planet for our growing population.

Malthus' initial concept of a population crunch arising from a mismatch between the growth of population and the availability of resources has inspired further considerations of factors that could lead human civilization into a "doomsday" that significantly reduces, completely eliminates, or otherwise transforms our population. A more recent version known as the "doomsday argument" implies that we might be much closer to the end of civilization than to its beginning due to the probabilistic implications of this growth (Carter and McCrea 1983; Bostrom 2013) or more broadly from the Copernican principle (Gott 1993). While the merits of this particular line of reasoning for the doomsday argument remain under debate (e.g., Korb and Oliver 1998; Bostrom 1999; 2013; Olum 2002; Bostrom and Ćirković 2003; Monton 2003), the general concept of a "doomsday" (in this case, a population catastrophe) arising from fundamental limits to growth remains a possible and plausible trajectory for human civilization.

Sebastian von Hoerner was a German astrophysicist who spent much of his career at the National Radio Astronomy Observatory in Green Bank, West Virginia. His research included work on stellar dynamics and radio antenna design, but his time at Green Bank, working alongside Frank Drake, allowed him to also become one of the pioneers in the search for extraterrestrial intelligence (SETI) (von Hoerner 1961; 1962). In thinking about SETI, von Hoerner also turned to consider projections of human civilization's growth across the planet and into space, considering the environmental and technological factors that place limits to our growth. His paper "Population Explosion and Interstellar Expansion" (von Hoerner 1975, hereafter VH75) considers the limits of food, energy, and territory in the future growth of our population, with calculations of the expected time until "doomsday," when he predicted human civilization would reach these particular limits to growth.

In this paper, we update the calculations of VH75 with the advantage of hindsight and the availability of newer data on human population growth and global energy consumption. We refine the time horizons of VH75 to show that any expected "doomsday" tends to be pushed further into the future, typically by centuries or more. We also discuss the immediate concerns of anthropogenic climate change, which VH75 did not fully consider, as well as the longer-term problem of direct thermal heating of the planet. We argue that these thermodynamic limits imply that humanity cannot sustain its present growth rate in energy use and possible evolution into a colonizing, spacefaring civilization without significant negative consequences to its supporting terrestrial environment. This suggests that the adoption of sustainable development practices may be a requirement for human civilization, and possibly for any technological civilization, to avoid a population- or energy-related "doomsday." In this work we will keep the term "doomsday," but place it in quotations and treat its connotations liberally. While it had strict meaning in VH75, this term more realistically refers to a time limit by which transformative changes in population growth, energy use, and/or some other structuring of civilization are required to ensure survival. It may or may not mean "doom" in an irreversible, apocalyptic sense, depending on how we respond to the limits and timescale it implies.

## 2. Limits to Population Growth

*2.1. Population Projections, Then and Now*

In general, the rate of change of a population with respect to time, $dN/dt$, depends on the population itself, $N$, and the relative growth rate of that population per unit time, $\gamma$, through the following differential equation:

$$\frac{dN}{dt} = \gamma N, \qquad (1)$$

where it is assumed $\gamma$ can be a function of $N$, $t$, and/or other factors. Around the time of its publication, VH75 (with data from Basler 1971) estimated that the human population was about 3.7 billion and increasing by about 74 million people per year. This corresponds to a yearly growth rate of ≈2.0% yr$^{-1}$. If $\gamma$ is constant, the solution to Eq. (1) yields exponential growth in $N$ with respect to $t$, with a doubling timescale $\ln(2)/\gamma$ ≈35 years.

Certainly, exponential growth of any population is eventually unsustainable on a finite planet with finite resources and should be cause for concern. However, von Foerster, Mora, & Amiot (1960) found that historic population trends from antiquity through the mid-20th century indicated a population dependence of the growth rate $\gamma$, i.e., $\gamma(N) = \gamma_0 N$, with a fiducial growth rate $\gamma_0$= 5.0 x $10^{-12}$ $yr^{-1}$ This provides a different solution to Eq. (1), specifically one where the population $N$ becomes a singularity in a finite timescale. VH75 calculated this timescale to be ≈54 years from 1972, corresponding to a population "doomsday" of 2026.

With less than a decade left until this predicted, population-induced catastrophe, where do we stand as a species? The population has indeed increased markedly since the 1970s, but as of this writing, there are about 7.5 billion people on the planet (United Nations, Department of Economic and Social Affairs, Population Division 2017c), well shy of the ≈25 billion that one would expect if the original predictions of VH75 and sources therein were accurate. Evidently, the global relative growth rate has declined since the 1970s, slowing the increase in population since that time. In fact, most robust predictions show that this trend will continue and $\gamma$ should continue to decline well into the mid-21st century (Raftery 2012; United Nations, Department of Economic and Social Affairs, Population Division 2017a, 2017b), likely following trends in economically and socially-driven declining fertility in both the developed and developing world (see, e.g., United Nations, Department of Economic and Social Affairs, Population Division 2017d for a review). Within a 95% confidence interval, these models generally predict that the human population should reach their maxima of ≈10–20 billion around 2050 to 2100 or even later, before stabilizing or even declining.

It appears that the alarming population-induced catastrophe predicted in the mid-20th century will not actually come to pass. These admonitions may have been premature; populations may go through drastic changes in average fertility and family size as living conditions and education improves with time (United Nations, Department of Economic and Social Affairs, Population Division 2017d), which affects the global behavior of $\gamma$. However, Warren (2015) notes that projections have usually underpredicted actual populations, attributing the problem to the severe sensitivity of exponential functions to uncertainties in $\gamma$. Simply put, "renegade" subpopulations with even slightly higher growth rates can overwhelm the net population in only a few generations and undercut the predictive power of any population model.

With that in mind, the more "extreme" models of Raftery (2012) may be considered. Varying their suite of projections by +0.5 child per mother increases the net population to almost 17 billion by 2100, and increasing further until ~2400, when the predicted $\gamma$ finally decreases to ≈0. Other models offer more population growth; if there are virtually no changes in global fertility (possibly through an extreme influence of "renegade" populations), then the population by 2100 may reach about ≈25 billion or more and continue to increase into the 22$^{nd}$ century. An important issue moving forward to a potential headcount of 10–20 billion or more in the next hundred years may be growing enough food to feed everyone. "Doomsday" in principle may result from a fundamental limit of Earth's agricultural capacity (Denkenberger and Pearce 2014).

*2.2. Agricultural Limitations*

A first order estimate of the number of people that the planet can agriculturally sustain can be easily made. VH75 made one by reasoning that, with ≈3.7 kcal g$^{-1}$ available from wheat or similar crops, an average human consumption of 2500 kcal day$^{-1}$, and good agricultural yields of ≈3 tons hectare$^{-1}$, the planet could sustain a maximum population of ≈180 billion people, if all land masses on Earth were cultivated. That limit would be reached around 2025, about a year before the initially predicted population singularity. Although 180 billion is well over the updated population estimates of Section 2.1, a more accurate estimate can be obtained with current agricultural data (and the consideration that we probably don't want to turn all of Earth's real estate into farmland anyhow).

First, while staple crops like wheat, soy, corn, and rice, can range in caloric density from ≈1–4 kcal g$^{-1}$, let us stay with 3.7 kcal g$^{-1}$ for comparison with VH75, and similarly maintain an average human consumption of 2500 kcal day$^{-1}$. Let us also assume that this strictly vegetarian diet with no waste is preferred for maximizing agricultural efficiency and population size. According to data from the USDA database (Earth Policy Institute from U.S. Department of Agriculture, Production, Supply, & Distribution 2013), between 1970–1975, the worldwide average yield for grains was actually between 1.63–1.82 tons/hectare, which would drop VH75's original population limit to about 100 billion. However, both world grain yield (tons hectare$^{-1}$) and production (tons) have been rising between 1950–2010. If the 2012-era yields and production values are used instead, along with the total amount of *currently arable* land – land that is being used to grow crops now, or about 1.41 billion hectares (FAO 2017) – then the maximum population is ≈17 billion. This increases to ≈60 billion if we use all possible agricultural land on Earth (all land that is potentially cultivable; FAO 2017), and to ≈180 billion if all land on the planet is used. This brings us back to the original prediction.

But let us now suppose that production and yields continue to rise into the future. In fact, the trends have been remarkably linear from 1950–2010; USDA data (Earth Policy Institute from U.S. Department of Agriculture, Production, Supply, & Distribution 2013) show that the latter has been growing by about 0.038 tons hectare$^{-1}$ yr$^{-1}$, and the former by about 26.8 tons yr$^{-1}$. If these trends continue, then the amount of food available each year increases, as does the maximum population that it can sustain. With these data, and the total amount of *currently arable* land, the maximum population is ≈40 billion. This increases to ≈130 billion if we use all possible agricultural land on Earth, and to ≈400 billion if all land on the planet is used. These values are greater than all UN (2017) projections, again with the assumption of vegetarian diets, high caloric yields, and no food waste.

Figure 1 summarizes our results. Dotted black curves show the discussed models within the 95% confidence interval, solid black curves are variants with ±0.5 child per mother. The solid gray lines are the most "extreme" models with constant fertility rates (top) and constant fertility and mortality rates (bottom). Meanwhile, the blue and red curves show maximum population based on the preceding discussion, based using all arable land and possible agricultural land, respectively, to grow crops. The solid colored areas within the curves show the regions between maximum populations allowed with a constant crop yield (as of 2015), and for their projected increases based on the historic data. From this perspective, should overall agricultural production yields continue to improve, then the rate of food production will likely outpace population growth. This is true in all but the most extreme of predictions, where it would seem "doomsday" would occur around 2075–2100. In order to circumvent this, (and more pragmatically account for some degree of dietary omnivorism and inefficiencies in food processing and distribution), additional land may be required by the end of the century, and this in principle can be accomplished. The red curves in Figure 1 are still well above the most "reasonable" population projections, so unlike the predictions of VH75, most of the planet should stay safe from agricultural development.

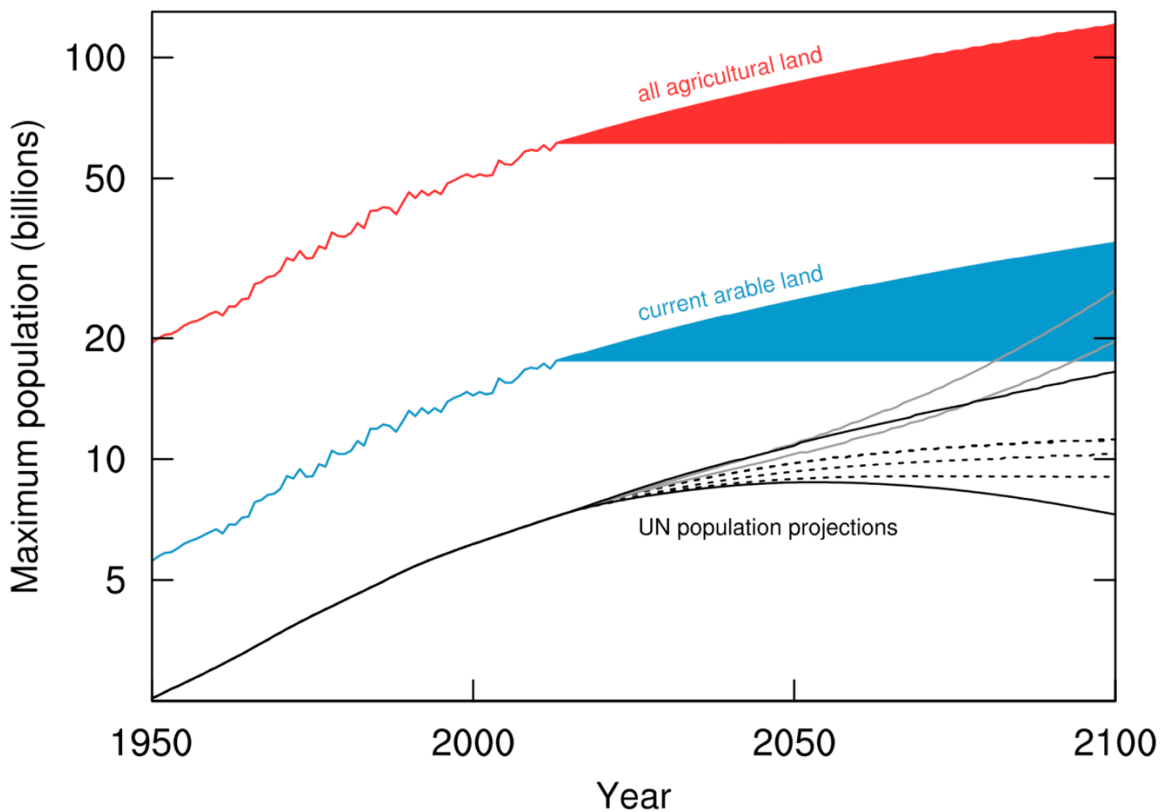

*Figure 1.* *Population projections (from Raftery 2012; United Nations, Department of Economic and Social Affairs, Population Division 2017a, 2017b), compared to maximum agriculturally sustainable populations. Dotted black curves are models within the 95% confidence interval, solid black curves are variants with ±0.5 child per mother. Solid gray curves are extreme models with constant fertility rates (top) and constant fertility and mortality rates (bottom). Blue and red curves show maximum population levels based using all arable land and all possible agricultural land, respectively, to grow crops (based on yield and production data from USDA 2013); see text for details.*

However, feeding this number of people will require high agricultural production capacity and efficiency. Increased production and efficiency occurs because of increased reliance on large-scale, energy-intensive agricultural practices (Diamond 1997). Energy use through both nonrenewable and renewable resources can have strong effects on Earth's climate, which may have disastrous effects on the planet's habitability for human beings. This provides a different avenue through which to estimate a potential "doomsday."

**3. Energy Use and Effects on Climate**

*3.1. Greenhouse Gas Emissions*

Anthropogenic climate change from the emission of greenhouse gases (GHGs) is a clear reality (IPCC 2013), though it was not as widely recognized in the 1970s. VH75 consequently does not mention it. Our large-scale use of energy through fossil fuels has had an impact on Earth's climate system; to date raising the mean temperature by ≈1K over its preindustrial average (USGCRP 2017), and leading to a possible continued warming by 2–8K by 2100 (IPCC 2013), and perhaps upwards of 10K or more further into the future (Stocker et al. 2013). This range reflects the uncertainty of various feedback mechanisms, such as additional greenhouse gas emissions from melting permafrost, reduction of Earth's albedo from melting polar icecaps, and effects from changes in cloud generation and cover (IPCC 2013 and references therein). Anthropogenic climate change appears unlikely to trigger a runaway greenhouse, which would render the planet uninhabitable (Goldblatt and Watson 2012; Ramirez et al. 2014); however, other factors such as heat stress limits (Sherwood and Huber 2010) or "double catastrophe" scenarios (e.g., Baum et al. 2013) could constitute existential risk for human civilization. Like agriculture, it would seem the hope is that using our resources (energy) will help us work around the problem by building "smart cities," engineering more resilient crops, managing our lives around drought, hurricane, and wildfire zones, and perhaps engaging in geoengineering.

One limit to consider may be what temperatures humans can physically endure. Sherwood & Huber (2010) estimate that the heat stress tolerance of mammals provides a limit to global mean temperature increases of ≈7K, primarily for the tropics. On a more planet-wide basis, the limit may be ≈12K. In terms of human survivability alone (ignoring what may be immense ecological effects to focus on one species' climatological adaptability) we could potentially experience a climate "doomsday"[5] due to ≈12K of warming by ~2300, if the upper end of predictions are correct and if national governments continue to fail in developing effective mitigation strategies (though Stocker et al. 2013 warn that this long-term forecasting is difficult to accurately accomplish).

---

[5] Migration of human populations from tropical regions to temperate and subarctic latitudes at temperature changes of ≈7K may be possible, but a mass migration of billions of people would likely cause rapid and dramatic social shifts and ecosystem changes, which would thus constitute a transformative "doomsday" of current population patterns and geopolitics.

*3.2. Direct Heating*

One cannot simply throw energy at a problem like agricultural production, or powering civilization in general, without inducing some change in the environment. In the broadest sense, our present situation amounts to changing the thermal emissivity of the atmosphere through our carbon emissions. However, an additional, infrequently mentioned source of climate change must also be considered, i.e. our thermal emissions into the climate system. Elementary thermodynamics and energy balance dictates that energy cannot be created nor destroyed. If we consider a "steady state" scenario wherein we assume most energy acquired is not stored over very long periods of time (see, e.g., Wright 2014b), then the energy we use is inevitably released as thermal infrared energy into the biosphere and radiated into space. It is not an issue of energy efficiency, but a matter applying the conservation of energy over the entire Earth system.

Remarkably, VH75 predicted this thermodynamic climate change (while missing the more immediate greenhouse gas problem), speculating that a maximum allowed change in temperature would be about ≈1K, which corresponds to worldwide power use of 2.1 x $10^{15}$ W. Provided a 7% growth rate in energy use estimated in western economies at the time (Basler 1971), VH75 estimated that this limit would be reached in ≈2050. Now that we have an idea of what temperature change we need to avoid to ensure survival (≈12K) and updated data on world energy use between 1850-1995 (Grubler 1998) and between 1995–2015 (OECD 2017), we can make a better estimate of thermodynamic "doomsday." Figure 2 shows the historic power use (in W) from 1850-2015. First, given the logarithmic scale, it can be seen that either the full range or some subset of the 1850-2015 data can be approximated by an exponential function:

$$P = P_0 e^{\gamma t} \qquad (2)$$

(the solution to Eq. (1) if $N$ were replaced by $P$ and $\gamma$ is constant), for a fiducial power $P_0$, a power use $P$ at a time $t$ after the corresponding fiducial time, a and relative growth rate $\gamma$. Overall, it would appear $\gamma$ ≈2.6% $yr^{-1}$. A more accurate fit for the timeframe of 1975–2015 can be obtained (see insert of Figure 2); this "current" growth rate is about 2.0% $yr^{-1}$. Either value is much less than the value of 7% $yr^{-1}$ VH75 used; that publication concentrated solely on western economic activity during a period of particularly strong economic growth.

Figure 3 takes these three representative values for $\gamma$, 2% $yr^{-1}$, 2.6% $yr^{-1}$, and 7% $yr^{-1}$, and projects these hypothetical trends into the future. In addition, we plot a toy logistic model for an arbitrary maximum power of 6.34 x $10^{13}$ W (2000 EJ $yr^{-1}$) and midpoint of 2050. While this function is also an adequate fit to the historical data, it is not *a priori* motivated by historical or physical reasons. It is meant as an illustrative example for later discussion. With these projections in mind we can include direct heating as a consequence of increasing power use $P$ taking two different forms: 1) anthropogenic forcing through the use of nonsolar energy, and 2) adjusted solar forcing through the use of solar energy.

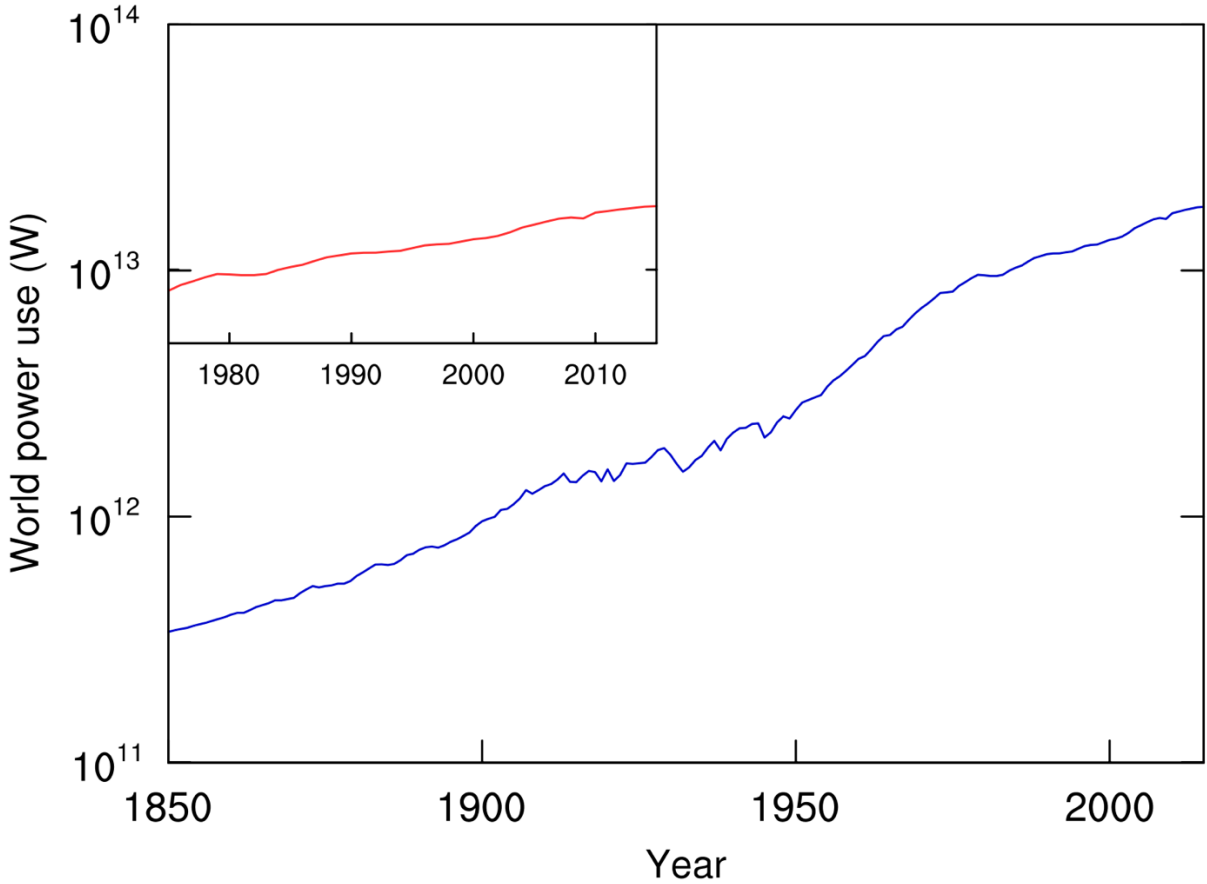

*Figure 2. World energy use between 1850-2015, from Grubler (1998) and OECD (2017). Insert: subset of data between 1975 and 2015.*

*3.2.1 Direct Heating with Anthropogenic Forcing*

To illustrate the scope of the problem that occurs, a basic energy balance equation with the Stefan-Boltzmann law and anthropogenic forcing can be established, where the power radiated by Earth at any time, $P_\oplus$, must equal the combination of absorbed solar radiation and the power used by human civilization:

$$P_\oplus = 4\pi\sigma\epsilon R_\oplus^2 T_\oplus^4 = (1-a)P_\odot + P, \qquad (3)$$

where $P_\odot$ is the integrated amount of solar power incident over the Earth ($\approx 0.25\, L_\odot R_\oplus^2 d_\oplus^{-2}$; $L_\odot$ is the sun's luminosity, and $d_\oplus$ is the distance from the Earth to the sun, and $R_\oplus$ is Earth's radius). $T_\oplus$ is Earth's global mean temperature, $\sigma$ is the Stefan-Boltzmann constant, and $\epsilon$ and $a$ are Earth's overall emissivity and albedo, respectively. This is essentially what VH75 did to establish "doomsday" via climate change. The only difference here is that we will be dealing with larger temperature increases, and will allow room for terrestrial albedo and emissivity to change.

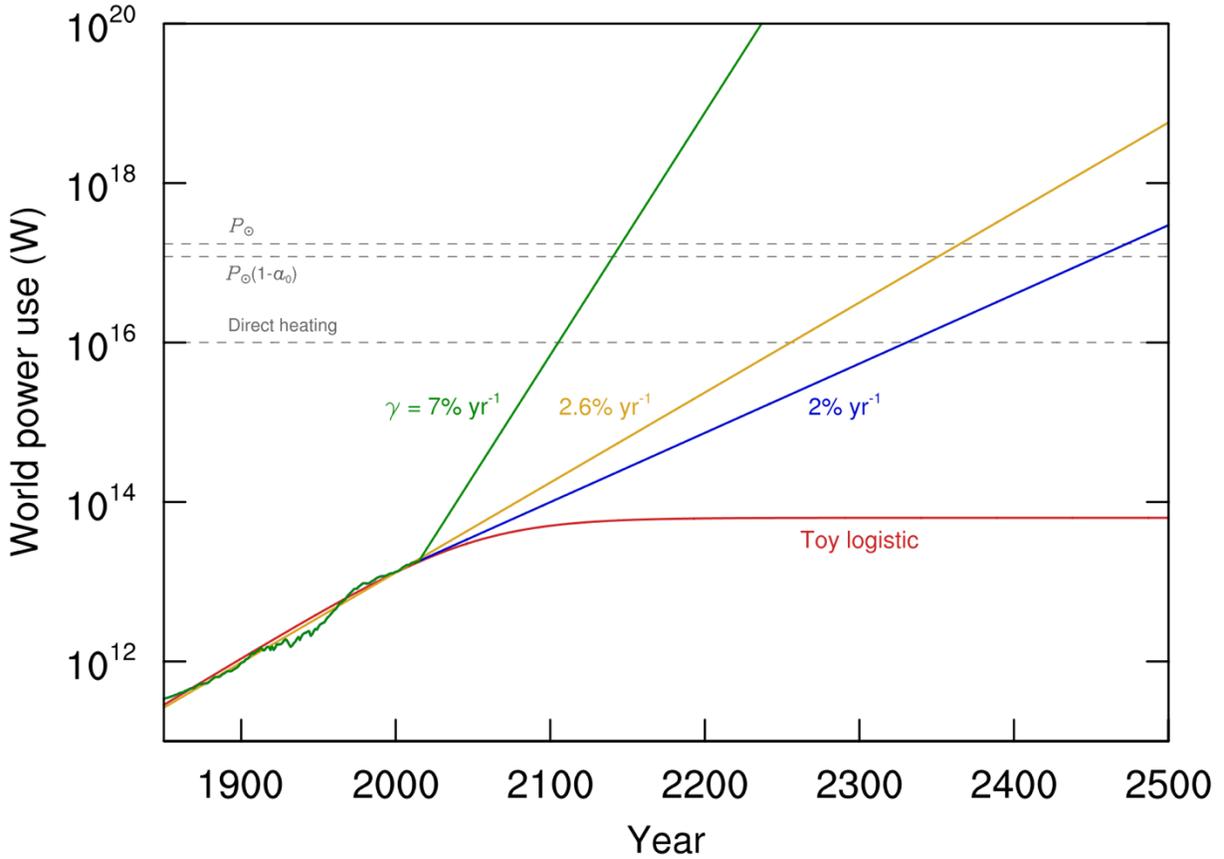

*Figure 3.* Example projections of the world power use data from Figure 2 (green; 1850-2015). Future tracks for growth rates of $\gamma$ = 0.07 yr$^{-1}$ (green), 0.026 yr$^{-1}$ (yellow), and 0.02 yr$^{-1}$ (blue) are shown, as well as an arbitrary toy logistic model (red) for comparison, with a maximum power 6.34 x 10$^{13}$ W (i.e., 2000 EJ yr$^{-1}$) and midpoint of 2050. Dashed gray horizontal lines are various power use thresholds: 10$^{16}$ W, about where direct heating becomes an important contributor to climate change (bottom); 1.2 x 10$^{17}$ W, approximately the power from the sun absorbed by Earth currently (middle); 1.73 x 10$^{17}$ W, approximately the power from the sun incident on the top of the atmosphere (top).

In this subsection, it is assumed the source of $P$ is strictly nonsolar: fossil fuels, nuclear, biomass, or a combination thereof. This is an admittedly simple model, but it suffices for our purposes – a direct comparison to VH75 given newer energy consumption trends.

At the present time, $P\left((1-a)P_\odot\right)^{-1} \sim \mathcal{O}(10^{-4})$, so direct heating is negligible compared to the impacts of GHGs. However, Figure 3 shows several possible trajectories where this may not always be the case. To establish a rough rule of thumb, let us say that direct heating is no longer negligible when $P\left((1-a)P_\odot\right)^{-1} \sim \mathcal{O}(10^{-1})$, or to pick a value, $P \approx 10^{16}$ W. This limit is seen as the bottom horizontal gray line in Figure 3. Therefore, direct heating becomes an important contributor to climate change by ≈2330, 2260, and 2100 for $\gamma$ = 2% yr$^{-1}$, 2.6% yr$^{-1}$, and 7% yr$^{-1}$, respectively. As to be expected, the logistic projection for the arbitrary parameters it was given never reaches this limit.

This doesn't constitute "doomsday," however. That will depend on when $T_\oplus$ increases past an acceptable limit. Combining Eq. (2) and Eq. (3), we can write the equation:

$$t = \frac{1}{\gamma}\ln\left(\frac{P}{P_0}\right) = \frac{1}{\gamma}\ln\left(\frac{P_\odot(1-a_0)}{P_0}\left[\frac{\epsilon}{\epsilon_0}\left(1+\frac{\Delta T_\oplus}{T_\oplus}\right)^4 - \frac{(1-a)}{(1-a_0)}\right]\right). \qquad (4)$$

Here, $\Delta T_\oplus/T_\oplus$ is the limiting fractional change in Earth's global mean temperature, $t$ is the time to that change, $\epsilon$ and $\epsilon_0$ are the final and initial values for Earth's emissivity during $t$, and $a$ and $a_0$ are the final and initial values for Earth's albedo during $t$. In principle, $a$ and $\epsilon$ are dependent on temperature and/or human activity. Generally, planet-wide albedo depends on factors such as cloud-cover fraction, cloudy-sky planetary albedo, and clear-sky planetary albedo over various surfaces, which can be affected by temperature directly or indirectly, e.g., changing weather patterns and ice melt. Here, we have a fiducial $a_0 = 0.31$ calculated from these effects (Trenberth 2009). If albedo is allowed to change, as it probably would with the temperature changes predicted (IPCC 2013), or emissivity (which continues to change as the atmosphere circulates the excess carbon dioxide we have pumped into it), then the corresponding power limit decreases, and the time to reach the new limit decreases.

Trying to delay catastrophic warming in this context may require immediately controlling net emissivity and albedo. Depending on the energy source, a substantial development of carbon capture and/or sequestering technologies may be needed to manage emissivity. Geoengineering provides a mechanism for changing or maintaining albedo, through large-scale efforts at altering planetary energy balance known as "solar radiation management" (SRM) (Caldeira et al. 2013). SRM proposals usually focus on the use of stratospheric aerosol to reflect incoming sunlight (Rasch et al. 2008), although similar reduction of incoming stellar energy could also be accomplished by orbiting mirrors or lenses (Angel 2006). Geoengineering solutions of this type would succeed in lowering the total amount of power absorbed by Earth, but it may necessitate a proportional increase in $P$ (through higher growth rates) for to accomplish more "desirable" human activities due to the power demands of maintaining such a geoengineering program.

Such strategies may help mitigate the anthropogenic carbon problem, but they cannot alleviate the thermodynamic limitations imposed by a growing energy use on a finite planet. Keeping $a$ and $\epsilon$ constant in Eq. (4) still results in a power limit that depends on temperature. Moreover, $P$ itself is independent of $\gamma$ and a limit for $\Delta T_\oplus/T_\oplus$ will always correspond to a set power, regardless of how long it takes a civilization to get there. Stalling growth in our capacity for energy use would affect the value of $\gamma$ and delay the reaching of power limits, but those limits remain. For a maximum warming of 12K and no changes to $a$ and $\epsilon$, that limit is ≈2.2 x 10$^{16}$ W, or about 12% of the total solar power incident at the top of the atmosphere. For comparison, a maximum warming of 7K drops those limits to ≈1.2 x 10$^{16}$ W (7% incident solar power). While energy use and the underlying economic growth that encourages it may feed a population faster than it can grow and allow room for other activity, there is no circumventing the first law of thermodynamics.

If we prevent changes in albedo and emissivity and maintain $\gamma$ at 2% yr$^{-1}$, we calculate that a mean warming by 12K will be reached by 2370. Using $\gamma$=2.6 yr$^{-1}$% decreases that horizon to 2290. For comparison to VH75, we also compute $t$ for $\gamma$=7% yr$^{-1}$ and find that "doomsday" would reached much earlier, by ≈2120 in this "high growth" case. For each of these cases, these "doomsday" times are at most only a few decades after when direct heating becomes important (where the curves intersect the established bottom horizontal line in Figure 3).

*3.2.1 Direct Heating with Adjusted Solar Forcing*

Would explicitly selecting another form of energy help? There is evidence to suspect that nonrenewable resource extraction may follow a logistic curve (see, e.g. Marder, Patzek, and Tinker 2016 for a review), and that within decades fossil fuel use will decline in favor of increasingly inexpensive carbon neutral alternatives like solar energy. Relying exclusively on solar power removes the direct anthropogenic climate forcing through a non-solar power source $P$ in Eq. (4). However, it makes up for it through planetary albedo changes and the thermodynamic consequences of taking that solar energy and, by using it, releasing it as heat into the biosphere (assuming again the steady-state scenario).

To first order, we can replace Eq. (3) with:

$$P_\oplus = 4\pi\sigma\epsilon R_\oplus^2 T_\oplus^4 = (1-f)(1-a)P_\odot + f(1-a_p)P_\odot \quad , \quad (5)$$

where the power collected by humanity is $P = f(1-a_p)P_\odot$, for a planetary coverage fraction $f$ of a "solar panel"- type technology (more generally some surface that collects solar energy) of albedo $a_p$. Instead of Eq. (4), we can then write:

$$t = \frac{1}{\gamma}\ln\left(\frac{P}{P_0}\right) = \frac{1}{\gamma}\ln\left(\left(\frac{1-a_p}{a-a_p}\right)\frac{P_\odot(1-a_0)}{P_0}\left[\frac{\epsilon}{\epsilon_0}\left(1+\frac{\Delta T_\oplus}{T_\oplus}\right)^4 - \frac{(1-a)}{(1-a_0)}\right]\right) . \quad (6)$$

Here, we have assumed $a_p$ can vary from 0 for optimal power collection[6] to any particular value to account for cloud cover or any other factors. The only difference between Eq. (4) and (6) is an increase in $t$ by an amount $\gamma^{-1}\ln((1-a_p)(a-a_p)^{-1})$.

Increasing our power use (by increasing the covering fraction $f$) may change both $a$ and $\Delta T_\oplus/T_\oplus$ simultaneously, and any decreases in $a$ and $\epsilon$ would decrease $t$, but again let us consider scenarios where the albedo of the uncovered Earth and emissivity is constant. With $\gamma = 2$% yr$^{-1}$, 2.6% yr$^{-1}$, and 7% yr$^{-1}$, and $a_p$=0 (perhaps an approximation for low-albedo absorbers that may move themselves to preferentially cloudless areas), "doomsday" occurs around 2430, 2330, and 2130, respectively. All these "doomsdays" are up to several decades after their counterparts of Section 3.2.1. Additionally, a temperature change of 12K corresponds to a power use of ≈6.9 x $10^{16}$ W, or about 40% of the total incoming solar radiation (and because $a_p$=0 this corresponds to a covering fraction f of ≈40% as well). Section 4.2 below explores the implications of higher values for $a_p$.

---

[6] Realistically, Rayleigh scattering and aerosol in the atmosphere may make the minimum $a_p$ closer to 0.06, but given the other uncertainties in our estimates of "doomsday" times, this effect is relatively small.

Of course, more accurate results require a more accurate analysis. Here, we have ignored second-order effects and feedback between the mass of solar collectors and the planet on which they rest. In addition, the chief factor in the last two subsections has been $\gamma$. While the historical record shows remarkable consistency, there is scatter in the data from a panoply of historical events, economic and social factors, technological innovation, etc. A constant $\gamma$ may be adequate now, but it may not hold up centuries in the future. For instance, it is conceivable that a logistic function like the one plotted in Figure 3 is more accurate, with a maximum energy use set around $\mathcal{O}(10^{14})$ W. From the above results, "only" around ~4-5K of warming may result (from GHGs alone), which may be ecologically problematic, but wouldn't necessarily constitute "doomsday" as it has been discussed here.

Thus, the "good" news is that, as long as we tackle carbon emissions and/or regulate Earth's emissivity and albedo, and maintain or decrease $\gamma$ (possibly by controlling our economic growth), we should have some time to anticipate direct thermodynamic heating from our energy use. The bad news is that, barring some new as of yet-undiscovered physics, our sensitivity to our waste heat sets a possible limit to our energy ambitions on Earth, at $\mathcal{O}(10^{16})$ W. At that point, the planet may be too warm for humans.

## 4. Implications for Interstellar Expansion

*4.1. Population-Driven Colonization*

If our climate situation is truly dire and/or our thirst for energy unquenchable, is it possible to take some fraction of our civilization or technology elsewhere, to colonize[7] other worlds? Starting a colony elsewhere does ensure our species' survival even if we self-destruct on Earth. Thus "doomsday" in this section more likely means a relatively sudden and necessary transition from one paradigm to another in terms of growth, expansion, or "lifestyle," and is less likely to mean a species-ending catastrophe. Hereafter, "doomsday" dates may be considered "doom" for what our species is doing at the time, and roughly marks a transition to a new paradigm of population or energy growth.

VH75 examined this possibility in terms of population pressure, the key issue that drove the bulk of the original text. A simple calculation reveals that with the then-documented ≈2% growth rate of the population, a colonization front would need to expand at an equivalent rate of 2% per year (in the relative growth in colonized volume, $\dot{V}/V$). At some point, the colonization front would have to approach the speed of light to maintain this expansion (the "light cage limit," c.f. McInnes 2002), and this corresponds to the maximum physically colonizable volume of space with all settled worlds therein. According to the original calculations, this provides a maximum radius to our potential "empire" of ≈50 pc, which would be reached by ≈2500[8].

---

[7] In this paper, we will consider "colonizing" to be synonymous with "settling." We use this word without political connotations.

Given the results of Section 2.1, one may expect that contemporary, decreased values of the population growth rate $\gamma$ would translate to a larger galactic footprint, achieved over a longer timescale. Indeed, with the current projections, taking the current population and applying $\gamma$ = 0.87–1.23% yr$^{-1}$ yields expansion distances of ≈70–100 pc. Assuming an average stellar density in the solar neighborhood of ≈0.14 stars pc$^{-3}$ and number of sufficiently Earthlike worlds ≈0.2 from recent Kepler Space Telescope data (Cassan et al 2012; Petigura et al. 2013), we may then hope to occupy $\mathcal{O}(10^5)$ worlds in the next 800–1400 years. "Doomsday" in this case, when the colonization wave speed $v = c$, would occur between 2900–3400.

If instead we use the population predictions of 2100 and their predicted growth rates as our starting point, the "doomsday" horizon is pushed even further into the future. With highly divergent growth rates of $\gamma$ = -0.1–2% from the wide variety of models considered, "doomsday" ranges wildly from ≈2500 in the most extreme, highest growth, highest fertility cases to never in models with declining $\gamma$. Most "reasonable" models, i.e. ones within a 95% confidence interval (Raftery 2012), predict colonization "doomsday" in the relative far future, $\mathcal{O}(10^4)$ years from now. The reward for such patience would be $\mathcal{O}(10^7)$ colonizable worlds over a colonizable volume of ≈500 –1000 pc in radius, but that is still a relatively small fraction of the entire Galaxy (≈30,000 pc in radius). Of course, if $\gamma$ is effectively 0 by that time as some models suggest, colonization is not driven by population pressure, and "simply" dividing up our civilization around the Galaxy can be achieved by any timescale and there is no "doomsday" (transition time) to consider.

*4.2. Energy-Driven Colonization*

Expanding beyond the bounds of Earth would also have the advantage of increasing humanity's access to energy. Once $\mathcal{O}(10^{16})$ W are obtained from either terrestrial sources or solar power, humanity fits the canonical classification of a Kardashev type-I (hereafter K1) civilization (Kardashev 1964; Sagan 1973a), i.e. a civilization that uses roughly a planet's worth of energy resources. As previously mentioned, this is about the same time that direct heating is an important factor in our planet's energy balance. Assuming the predictions of Section 3 are accurate, the Earth could be facing a potentially catastrophic increase in mean temperature by this time.

---

[8] As in the rest of the text, we are concerned here more with *physical* limits to expansion rather than *technical* limits. But in general, one may expect the maximum colonization radius to shrink from $R$ to $\beta R$ if an unforeseen technical limit imposes a colonization speed limit $v = \beta c$ for some fraction $\beta$, and for the time to that horizon to decrease by an amount $(3/\gamma)\ln\beta$. Certainly, including relativistic effects of propelling a colonization front forward at high speed into a civilization's energy budget would make this problem worse.

However, this temperature-related "doomsday" can be avoided by selecting nonzero values for $a_p$ in Eq. (6). More realistically higher values for $a_p$, e.g. for solar collectors that experience cloud cover during their operational lives, increase the time to "doomsday". While direct heating is no longer a problem in Eq. (6) as $a_p$ approaches $a$, we can consider a new "doomsday" at the point when the planetary coverage fraction $f = 1$ in the expression:

$$t = \frac{1}{\gamma}\ln\left(\frac{f(1-a_p)P_\odot}{P_0}\right) \quad (7).$$

If $\gamma$ is between 2% yr$^{-1}$ and 2.6% yr$^{-1}$, and $a_p = a = a_0$, we run out of room on the planet for human beings roughly between 2350 and 2460, though we would get about 1.2 x 10$^{17}$ W in the process. In comparison, if $\gamma = 7\%$ yr$^{-1}$, the corresponding time limit is 2140. This limit is plotted as the middle gray horizontal line in Figure 3.

Certainly, if this is our path, then we are probably not concerned with the planet's ecosystem. So what if we do not care about any temperature changes, either? A simple solution to avoiding heat stress effects of 12K of warming may be to have civilization live exclusively in climate-controlled habitats. In that case, more energy can be obtained by minimizing $a_p$ as much as possible in Eq. (7). With $f = 1$, having $a_p = 0$ provides the upper bounds on the energy obtained on/above/around Earth, conceivably by intercepting all of the solar power incident at the top of the atmosphere with orbiting collectors. Figure 3 shows this limit as the top gray horizontal line. That energy limit is just $P_\odot$ = 1.73 x 10$^{17}$ W, obtainable by 2470, 2370, and 2150 for $\gamma = 2\%$ yr$^{-1}$, 2.6% yr$^{-1}$, and 7% yr$^{-1}$, respectively.

From there, we could continue our growth in energy use and conceivably develop the capacity to harness the equivalent power of a star, $\mathcal{O}(10^{26})$ W. We would then be a Kardashev type-II civilization (hereafter K2), and later develop the capacity to use the equivalent of galaxy's worth of stars, $\mathcal{O}(10^{36})$ W as a Kardashev type-III civilization (hereafter K3). One possibility for a K2 civilization is constructing a "Dyson Swarm" (Dyson 1960), a large collection of solar power-collecting objects in orbit around a host star. VH75 originally predicted that, with a relative growth rate in energy use of $\gamma$ = 7% yr$^{-1}$, humanity would have to build such a structure by 2400. With $\gamma$ = 2.6% yr$^{-1}$ and our current energy use data, building a Dyson Swarm is pushed back to 3140, and taking over a Galaxy's worth of stars to 4030. Should "business as usual" continue as discussed in Section 3, beginning a transition to a K2 civilization would be a necessary one to continue our growth in energy consumption and we would necessarily become a totally spacefaring species (or at least relocate off of Earth) by ~2500. Decreasing $\gamma$ naturally delays this timeline.

## 5. Discussion

*5.1 Notes on "Doomsdays"*

It is worth pausing to reflect on the different "doomsday" timescales generated so far, and how they were obtained. Table 1 summarizes the main results of our calculations, and organizes the calculated "doomsdays" in approximate, updated, chronological order. All values are obtained from the preceding text, and for all energy-related "doomsdays," a growth rate of 2.6%, the projection based on the data in Figure 2, has been selected for comparison with VH75. Where appropriate, the comparison values for VH75 have been revised (e.g., for consistency in temperature limits) or newly generated for "doomsdays" not considered at the time.

**Table 1:** Summary of Results

| Event | "Doomsday" (VH75 or equivalent) | "Doomsday" (updated) | Notes/reasons for difference |
|---|---|---|---|
| Pessimistic agricultural production limits | 2025 | 2075 – 2100 | ● Estimate only uses currently arable land<br>● Estimate includes extreme population projections<br>● Pessimistic estimate is avoidable with additional land |
| Optimistic agricultural production limits | | never | ● Estimate considers all possible agricultural land<br>● Agricultural production will likely outpace population in all but extreme population projections |
| Population singularity | 2026 | never | ● $\gamma$ has declined since VH75<br>● Max. population likely ≈10-20 billion in next ≈100 years |
| Humanity becomes K1 civilization | 2080 | 2260 | ● Lower $\gamma$ (2.6%) compared to VH75 (7%)<br>● Calculation not performed in VH75, but inferred from given values |
| Climate Change (from GHGs) | n/a | ≈2300 | ● Set by ≈12K increase in global mean temperature<br>● Estimate from Stocker et al. (2013)<br>● Not considered in VH75 |
| Climate change (from direct heating) – nonsolar | 2090 | 2290 | ● Lower $\gamma$ (2.6%) compared to VH75 (7%)<br>● Set by ≈12K increase in global mean temperature<br>● Exclusively nonsolar power used<br>● VH75 calculation adjusted for consistent temperature limit |
| Climate change (from direct heating) – solar | | 2330 | ● Lower $\gamma$ (2.6%) compared to VH75 (7%)<br>● Set by ≈12K increase in global mean temperature<br>● Only solar power used |
| Earth covered by solar collectors | 2110-2120 | 2350 – 2370 | ● Only solar power used<br>● No limit set for increase in global mean temperature<br>● $0 \leq a_p \leq a$<br>● Calculation not performed in VH75, but inferred from given values |
| Limits to interstellar colonization (1) | ≈2500 | ≈2900 – 3400 | ● Projection with current population and estimated range in possible $\gamma$ |
| Humanity becomes K2 civilization | 2420 | 3140 | ● Lower $\gamma$ (2.6%) compared to VH75 (7%) |
| Humanity becomes K3 civilization | 2750 | 4030 | ● Calculation not performed in VH75, but inferred from given values<br>● Lower $\gamma$ (2.6%) compared to VH75 (7%)<br>● Not physically reasonable (cannot settle the Galaxy in ~1000 years) |
| Limits to interstellar colonization (2) | n/a | $\mathcal{O}(10^4)$ – never | ● Projection with modeled population of 2100 and estimated range in possible $\gamma$ |

As discussed in the preceding sections, several trends are apparent:

- **Growth rate $\gamma$ is crucial.** Much of this work and research cited amounts to pushing back the "doomsday" horizons back for many of the situations both VH75 and we examined. Several problems – the population singularity issue, optimistic forecasts for food production, and the light cage limit to interstellar expansion – may not actually ever occur. Others, like engineering benchmarks for Kardashev types, are delayed by hundreds of years. In these cases, the underlying physics and mathematics has not changed since VH75, but the empirical choice for growth rate $\gamma$ has. Indeed, the mathematics used in these estimates and in VH75 are uncomplicated and it should come as no surprise that $\gamma$ plays such a strong role. Rather, evidence for smaller growth rates now compared to VH75 is what is "new" here. Decreases in $\gamma$ (in terms of population and energy use) has prolonged the timescales to various "doomsday" thresholds by factors of 2 or more. Should $\gamma$ continue to decrease, the equations used in this text would certainly need to be modified, but clearly these "doomsdays" would continue to be pushed into the future.

- **"Doomsdays" are more interrelated than presented here.** Though it has been mentioned briefly thus far, it bears repeating that population, agriculture, energy consumption, and climate are interrelated. Thus each effect, while individually considered here, feed back on each other in complex, nonlinear ways. Our land use induces feedback effects on climate (e.g., Pielke Sr, Mahmood, & McAlpine 2016). Meanwhile, rising temperatures may have an adverse impact on agricultural productivity. Tack, Barkeley, & Nalley 2015 predict that agricultural yields for crops like wheat may suffer a ≈50% loss with a 5K increase in temperature. Population growth may also respond to food availability in ways not yet predicted. Between population/environmental feedback and efficiencies in production, distribution, crop variety, and omnivorous diets, the maximum realistic population we may sustain without using more land may be closer to ≈10 billion (c.f., Wilson, 2012). This population is reached rather soon according to most population predictions, around 2050. While more land may certainly be used, it would again require more (possibly nonrenewable) energy, and feed into a temperature feedback loop. This may have the cumulative effect of constraining "doomsday" closer to the present than discussed in Section 3.2. Ultimately, a holistic approach is needed to evaluate a comprehensive "doomsday;" groups like the *Planetary Boundaries Institute* (e.g., Steffen et al. (2015), Rockström et al. (2009)) are rigorously engaged in related work.

- **Climate change can occur from both GHGs and direct heating:** An updated calculation places "doomsday" from direct heating of the climate system up to several hundred years further back than VH75. This may give humanity some time to solve climate change by GHGs first (say, constraining emissivity and albedo by the time our power use becomes a non-negligible part of the planet's energy budget, or ~2250) before having to worry about unavoidable direct heating effects. While perhaps less pressing than our concerns for the immediate future (GHG effects by 2100, for instance), a more accurate analysis of the effects of climate change around this timescale will need to incorporate direct heating, a phenomenon that likely cannot be avoided unless $\gamma$ continues to decrease in the future (e.g., as expressed in the toy logistic function plotted in Figure 3). The response and feedback of emissivity and albedo to temperature changes as a results of GHGs and direct heating also needs to be assessed more rigorously than here. Ultimately, using a power consistent with K1 status, $\mathcal{O}$ ($10^{16}$) W, could induce a problematic temperature change (established here as $\Delta T_\oplus / T_\oplus \gtrsim 0.04$). The time to this change can be delayed by use of solar instead of nonsolar power, and this temperature problem could be avoided by carefully controlling solar collector albedo. However, an exponential growth in energy use presents a new planetary coverage problem soon thereafter.

- **When does the world end?:** If temperature changes are not considered a problem (possibly by living indoors and/or engineering our food, air, water, or ourselves) or we are careful in our solar power collection, our historic growth rate dictates that we may cover the planet with solar collectors by ~2400. This may be a necessary step on the way to becoming a K2 civilization and mark the end of the Earth before a Dyson Swarm must be constructed. If we are not a fully spacefaring civilization by then, we will need to be. This "doomsday" is therefore one for our lives on the planet, but possibly not the species itself. Alternatively, leaving the Earth alone to start building a Dyson Swarm soon may be an option. This could keep the planet's ecosystem unperturbed, at least until constructing the Dyson Swarm interferes with the solar energy it would otherwise intercept.

- **Can we become a K2 civilization?** In both VH75 and this work, population-driven interstellar expansion limits are reached around the same time as humanity approaches a K2-level civilization (in ~1000 years by our current estimates). There is nothing physically responsible for the concordance of these timescales; they depend on two separate growth rates (population and energy use) and stated thresholds for the Kardashev classification scheme. If the population growth projections from 2100 onward are more accurate, these timescales could be very different. If our population is constrained to the Solar System but maintains its thirst for energy, then ~1000 years could remarkably be enough time to build a Dyson Swarm (Armstrong and Sandbert 2013), though it would be a significant engineering challenge.

- **Can we become a K3 civilization?** Becoming a K3 civilization requires either an exotic power source or power from $\mathcal{O}(10^{10})$ stars. Similar to a K2 transition, a lethargically growing or declining population that pushes the expansion "doomsday" back indefinitely would still require some manner of technology "population" growth (interstellar expansion[9]) to become a K3 civilization. It is therefore possible that a consolidated, more fundamental "doomsday" may be considered for such a potentially spacefaring species, essentially the minimum of either expansion or energy consumption calculations. For our results here, it would appear that a lack of population pressure in the future may allow us to settle the Galaxy on whatever timescale we wish, but the ~1000 year transition from a K2 to a K3 civilization would require us to take over the Milky Way by 4030! Clearly that is not possible, and the growth rate of energy/technology "population" must continue to decline and push that timescale back by at least 3-4 orders of magnitude.

*5.2 Implications for Fermi's Paradox*

It is natural, then, to examine these findings through the lens of the Fermi paradox (Jones 1985; Hart 1975); i.e., "where is everybody?" The question is troubling given the disparity in timescales between the age of the Galaxy, $\mathcal{O}(10^{10})$ yr, and the amount of time it should take any sufficiently motivated species to visit all the habitable worlds within it, $\mathcal{O}(10^6$–$10^8)$ yr. VH75 considers several options, which can be confronted with our updated calculations and the hindsight of several decades of astrobiology research.

VH75 offers a few explanations that, while possible in principle, cannot be substantiated within the bounds of the original or this current work:

1) ***Life could just be extremely rare (or we are the first civilization in the Galaxy).*** This is certainly the simplest explanation, one that does not require the input of this work or

---

[9] E.g., "Von Neumann machines" and other self-replicating spacecraft (Freitas 1980)

VH75. This stance may be supported by the *Rare Earth Hypothesis* (Brownlee & Ward 1999), but see Kasting (2001) for a critique of this argument.

2) ***Extraterrestrial civilizations may experience a "change an interest" that prevents them from taking over the Galaxy.*** This explanation, like other sociological arguments, are impossible to demonstrate without any understanding of, or evidence for, underlying alien psychologies or social pressures.

3) ***Emerging civilizations like ours may self-destruct before, expanding into the cosmos.*** Indeed, if there is even a small probability of self-annihilating war happening in any particular time, the likelihood of destruction increases the longer a (contained) civilization lasts. Technological civilizations are subject to both global catastrophic and existential risks, some of which depend upon environmental factors but others of which depend upon sociological factors (Bostrom and Ćirković 2011).

4) ***More advanced civilizations may be out there, but we cannot see them.*** The last series of solutions that VH75 considers can be categorized as difficult-to-prove "others." These include the "zoo hypothesis" (humanity is kept in a sort of interstellar wildlife preserve), and the idea that hyper-advanced, K2–K3 civilizations do exist, but they don't take the form of something we can easily detect like Dyson swarms, and operate beyond our cognitive and physical capacity. None of these are possible to prove or disprove yet, but they indicate that our understanding of our present trajectory and the problems we face are myopic. The "mental horizon" argument of VH75 is similar to the "communication horizon" argument of Sagan (1973), who showed that, given the likely timescales of civilizations at different stages in their development, we are more likely to have more advanced civilizations (K2–K3) around than peer ($\lesssim$ K1) civilizations. This is evident in our calculations; the ~1000-year period between Kardashev types is much less than the geological or stellar timescales in which long-lived civilizations may operate. Consequently, we may be surrounded by advanced civilizations and their technology or interests may be so far beyond our comprehension we may not even notice them. In that case, we may not even have to worry about the kinds of "doomsdays" documented here! Alternatively, perhaps the Milky Way is undergoing a phase transition in habitability allowing civilizations to emerge now (Ćirković & Vukotić 2008). There are no shortage of possible solutions to the Fermi paradox (see Ćirković 2009 for a review).

VH75 offers one explanation in light of his alarmingly close doomsday predictions that may also be consistent with our updated calculations:

5) ***Civilizations need to stabilize themselves to avoid the self-inflicted catastrophes analyzed here.*** This "stabilization" solution to the Fermi paradox posits that a civilization surviving over the long term must protect itself from all external (astronomical and geological) and internal (population and activity-related) threats. VH75 concludes that surviving civilizations must carefully manage and curtail their growth and resource consumption to avoid the catastrophes he predicts for our own. Decades later, our follow-up has shown that our relative slowdown of growth in population and energy consumption have eliminated or delayed those catastrophes. Both approaches appear to favor the *sustainability solution* presented by Haqq-Misra & Baum (2009), where the only extant advanced civilizations in the Universe must prevent themselves from the kind of population growth, energy use, and interstellar expansion warned by VH75 that would cause them to self-destruct before we could detect them.

The "sustainability solution" has precedent in the literature, in a variety of contexts. In comparison, Newman & Sagan (1981) numerically model galactic colonization as a diffusive process to more accurately obtain estimates for the colonization timescale of the Milky Way, and find that the timescale can increase to $\mathcal{O}(10^{10})$ yr if the population growth rate decreases to ≈0. Vukotić & Ćirković (2012) find that, instead of homogenous expansion across the whole Milky Way, emerging species may end up confining themselves to isolated pockets of space of $\mathcal{O}(10^5-10^6)$ worlds. As we have seen here, energy use growth must drop accordingly, or else a civilization would need to expand faster than possible to obtain it. The growth to a K1 civilization and beyond may also slow for "sustainable" civilizations, and may be purposefully so to manage climate feedback by direct heating. A possible track for them might be something like the toy logistic model of Figure 3. Indeed, even if we were not concerned with these matters, our current track may in principle allow us to become a K2 civilization but would be impossible to sustain if we want to become a K3 civilization.

Frank & Sullivan (2014) and Frank et al. (2018) take a more rigorous look into emerging civilizations and climate feedback, and assert that sustainability may be an attractor state for civilizations, regardless of their development (see also Frank, Alberti, Kleidon 2017 for a more detailed discussion and classification of a civilization's entropic effect on its planet). This mirrors what we have found in Section 3, namely that it may be difficult to avoid becoming a K1 civilization without inducing a significant relative change in global mean temperature and/or covering much of the planet with solar collectors. Indeed, the K1 threshold of $10^{16}$ W is also our threshold for when direct heating from our power use becomes important in Earth's energy balance. This may mean that becoming a K1 civilization is simply unsustainable and no one else has wanted to do it or survived the process. Alternatively, perhaps we are the first civilization ever to try, or are among many peers in an aforementioned Galactic "phase transition." Of course, knowing as little as we do about other planets' civilizations (that is, absolutely nothing as of this writing), this is all speculative, but it is an interesting line of inquiry nonetheless.

*5.3 Implications for SETI*

If the sustainability solution is a viable explanation for the Fermi Paradox, then what would that suggest for observational searches for extraterrestrial intelligence (SETI), and what strategies may be recommended for optimizing success? There are no shortage of possible strategies (see, e.g. Korpela & Howard 2008 or Tarter 2001 for a review), but we may broadly categorize implications for two "branches" of SETI – communications (intentional or intercepted), or artifacts.

### 5.3.1 Communications SETI

VH75 calculates that if the fraction of stars in the Milky Way that have a civilization attempting to communicate is ~0.001-0.1, the communications timescale (the time to wait between messages) is approximately 2000–6000 years. Moreover, Frank & Sullivan (2016) find that, as long as the probability of getting a technological species in the habitable zone of a solar system is $\gtrsim 10^{-24}$, we should not be alone. But if it is that low, the timescale to wait would stretch to $\mathcal{O}(10^9)$ years. If there are long-lasting, sustainable sub-K1 civilizations that want to communicate, it is possible that we just haven't looked long enough. And while the "communication horizon" argument of Sagan (1973) would imply that civilizations would be far past our ability to communicate and even recognize them, the logic and mathematics of that argument changes if the sustainability solution is invoked. In short, the sustainability solution implies that civilizations do not make the transition to higher Kardashev types, and most civilizations in the Galaxy manage their growth to remain "at our level," slow their growth indefinitely, or self-destruct. Answering the question of how many are there (or survive the process of finding their equilibrium) would again require time, possibly on the order of thousands of years or more. In that case, we may be surrounded by intelligent life, and we just need to be patient. Guessing the right frequencies, methods of communication, and part of cosmos would take time and luck.

### 5.3.2 Artifact SETI

Energy use at the level of K2–K3 civilizations should be detectable at interstellar distances. Wright et al. (2014a, 2014b) established the viability of searching for K2 and K3 civilizations by the waste heat they must invariably produce (save some exotic physics we have yet to discover). The central idea is that a Dyson Swarm blocks some fraction of the emitted star or galaxy light, and reradiates that energy as waste heat; thus the altered spectrum of a star or galaxy immersed in a Dyson swarm should most obviously contain an artificial, Planckian component at mid-infrared wavelengths corresponding to the operational temperature of the solar collectors (~300K). Wright et al. (2014b) and Griffith et al. (2015) used mid-infrared WISE and Spitzer observations to search for Dyson swarms within the Milky Way (K2 civilizations) and Dyson Swarms covering a high fraction of stars in external galaxies (K3 civilizations). They found no evidence for such advanced activity, and nor did their predecessors (Carrigan 2009a, Carrigan 2009b, Jugaku & Nishimura 2004; Jugaku & Nishimura 2000; Jugaku et al. 1995; Jugaku & Nishimura 1991). This is what a sustainability solution would predict, though it is also conceivable that such advanced civilizations use some as of yet-undiscovered physics to bypass what we would expect in their thermal emissions.

Would K1 civilizations be observable? An emerging K1 civilization using some independent power source or some fraction of incident solar energy (c.f. Section 3.2) should also alter its mid-infrared planetary spectrum to incorporate the additional, presumably smooth blackbody component from their initial energy absorption and inevitable thermal radiation. In principle, extremely precise measurements of exoplanetary mid-IR spectral energy distributions (SEDs) could show such evidence of near-K1 activity beyond Earth. An excess of thermal emission compared to an otherwise similar but unoccupied planet, peaking at wavelengths corresponding to the operational temperature of a civilization's technology (presumably ~300K, ~10 $\mu$m) would be present. Kuhn & Berdyugina (2015) provide additional, related suggestions for detecting evidence of these civilizations.

However, certain "sustainable" civilizations may be the most difficult to detect. Civilizations that carefully control the coverage and the albdeos of their solar collectors may not experience artificial warming, nor have a demonstrable impact on their SEDs – they would just be intercepting, using, and reradiating energy that would be ordinarily absorbed and reradiated by their planets. From interstellar distances, these civilizations may be practically undetectable, at least to first order. But such (or any) sub-K1 to K1 civilization should be capable of sending remote exploratory spacecraft to nearby star systems (Rose and Wright 2004; Matloff 2006). The detection of any such extraterrestrial technology in either a functional or defunct form would represent conclusive evidence of the existence of extraterrestrial intelligent life – although locating a tiny spacecraft within the expanses of a stellar system would be a challenging feat to say the least (Haqq-Misra and Kopparapu 2012).

The sustainability solution provides another prediction for SETI. Between VH75 and this work, a civilization like ours can reasonably expect to expand its interstellar reach to $\mathcal{O}(10^5–10^6)$ worlds across tens of parsecs or more if driven by population growth. It could in principle leisurely visit all worlds in the Galaxy as long as it has a Hubble time's patience and is not driven by population growth (Newman & Sagan 1981). Additionally, Newman & Sagan (1981) expects ~$10^5$–$10^6$ worlds in such a civilization's slow diffusion are "linked by heritage." This clustering of occupied worlds is predicted elsewhere. Several models of interstellar dispersion mediated by varying cultural tendencies (Vukotić & Ćirković 2012, Carrigan 2012) find that civilizations can get hemmed in by frontier worlds that become unlikely to continue the colonization process; this could be interpreted here as attaining sustainability. Provided that we find such a civilization in a dense massive star cluster or in under a Galactic rotation timescale (to mitigate the effects of differential rotation and shear), It may then be possible that the search for one communicating or thermally augmented world reveals many more next door to it.

*5.4 The Ethics of Growth*

This discussion has focused on the physical limits to growth, but we cannot consider the prospect of future population/energy consumption growth into the Galaxy without also asking whether or not such an action *ought* to be done. The population projections discussed in Section 3 and shown in Figure 2 find that Earth could support a substantially larger number of people than are alive today; however, these uppermost projections of ~20 billion (by optimizing current arable land) or ~100 billion (by using all available land for agriculture) would require a much lower quality of life than enjoyed by many of us today. These extreme upper limits would require all people to adopt limited vegetarian diets on a planet transformed almost entirely for the production of food, likely requiring augmentation of crops (or ourselves) by bioengineering, with possibly extreme environmental consequences. Is such a future ethically preferable to one with fewer people, even if it could be a stepping stone towards a wholly different K2-K3 paradigm?

An influential version of this question was formulated by Derek Parfit (1984), which suggests that the value of (say) a world of 10 billion happy people could be equal to the value of a world of a vastly larger number of people living lives barely worth living. This "repugnant conclusion" at least requires us to give pause to the consideration that continued rapid growth in population or energy could tend toward such a future. We mention Parfit's idea briefly as a concluding thought, but this controversial population ethics question spans a substantial literature that remains beyond the scope of the present study. Similarly, our ethical obligations to our home planet, the effects of our energy use on its ecosystem, and the future we want for future generations are well beyond this cursory discussion, but are well worth considering and planning for.

While it may be "repugnant" to our senses to think of a distant future with 100 billion people with lives barely worth living, perhaps a future of 100 billion (or more) people ultimately living on other planets in other star systems *would* be a better future—even if the objective standard of living were lower than today. Civilizations that experience unsustainable growth, and attempt to adopt long term sustainable practices, will be forced to curb their rate of energy consumption that will likely require significant changes in lifestyle. Although we must be cautious about bestowing ethical obligations upon our future (and unborn) descendants, perhaps a reduction in lifestyle (and thus future energy consumption) would enable a much longer-term sustainable future for human civilization.

The authors wish to thank the anonymous referees for their input and suggestions that significantly improved this paper. This research did not receive any specific grant from funding agencies in the public, commercial, or not-for-profit sectors.